\documentclass[prl,final,twocolumn,showpacs,superscriptaddress,preprintnumbers,floatfix]{revtex4-1}

\usepackage{dynlearn}

\CMIndexedSymbol[xharpoon]{TSx}{X}
\CMIndexedSymbol[xharpoon]{tsx}{x}
\CMIndexedSymbol[xharpoon]{TSy}{Y}
\CMIndexedSymbol[xharpoon]{tsy}{y}
\CMIndexedSymbol[xharpoon]{TSz}{Z}
\CMIndexedSymbol[xharpoon]{tsz}{z}

\newcommand{\faircoin}{%
  \begin{cases}
    0 & \textrm{with probability } \half \\
    1 & \textrm{with probability } \half
  \end{cases}
}

\newcommand{\drawtea}{%
  \includegraphics{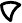}
}

\newcommand{\drawmia}{%
  \includegraphics{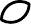}
}

\newcommand{\drawteb}{%
  \includegraphics{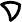}
}

\newcommand{\drawmib}{%
  \includegraphics{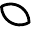}
}

\begin{document}

\def\ourTitle{%
  Information Flows?\\
  A Critique of Transfer Entropies
}

\def\ourAbstract{%
A central task in analyzing complex dynamics is to determine the loci of
information storage and the communication topology of information flows within
a system. Over the last decade and a half, diagnostics for the latter have come
to be dominated by the \emph{transfer entropy}. Via straightforward examples,
we show that it and a derivative quantity, the \emph{causation entropy}, do
not, in fact, quantify the flow of information. At one and the same time they
can overestimate flow or underestimate influence. We isolate why this is the
case and propose several avenues to alternate measures for information flow. We
also address an auxiliary consequence: The proliferation of networks as a
now-common theoretical model for large-scale systems, in concert with the use
of transfer-like entropies, has shoehorned dyadic relationships into our
structural interpretation of the organization and behavior of complex systems.
This interpretation thus fails to include the effects of \emph{polyadic}
dependencies. The net result is that much of the sophisticated organization of
complex systems may go undetected.
}

\def\ourKeywords{%
  stochastic process, transfer entropy, causation entropy, partial information decomposition, network science
}

\hypersetup{
  pdfauthor={Ryan G. James},
  pdftitle={\ourTitle},
  pdfsubject={\ourAbstract},
  pdfkeywords={\ourKeywords},
  pdfproducer={},
  pdfcreator={}
}

\author{Ryan G. James}
\email{rgjames@ucdavis.edu}
\affiliation{Complexity Sciences Center}
\affiliation{Physics Department}

\author{Nix Barnett}
\email{nix@math.ucdavis.edu}
\affiliation{Complexity Sciences Center}
\affiliation{Mathematics Department,\\
University of California at Davis, One Shields Avenue, Davis, CA 95616}

\author{James P. Crutchfield}
\email{chaos@ucdavis.edu}
\affiliation{Complexity Sciences Center}
\affiliation{Physics Department}
\affiliation{Mathematics Department,\\
University of California at Davis, One Shields Avenue, Davis, CA 95616}

\date{\today}
\bibliographystyle{unsrt}

\title{\ourTitle}

\begin{abstract}

\ourAbstract

\vspace{0.1in}

\end{abstract}


\preprint{\sfiwp{16-01-001}}
\preprint{\arxiv{1512.06479}}

\title{\ourTitle}
\date{\today}
\maketitle

\setstretch{1.1}

An important task in understanding a complex system is determining its
information dynamics and information architecture---what mechanisms generate
information, where is that information stored, and how is it transmitted within
a system? While this pursuit goes back perhaps as far as Shannon's foundational
work on communication~\cite{Shan48a}, in many ways it was
Kolmogorov~\cite{Kolm56a,Sina59,Orns89} who highlighted the transmission of
information from the micro- to the macroscales as central to the behavior of
complex systems. Later, Lin showed that ``information flow'' is key to
understanding network controllability~\cite{lin1974structural} and Shaw
speculated that such flows between information sources and sinks is a necessary descriptive framework for spatially extended chaotic systems---an alternative to narratives based on tracking energy flows~\cite[Sec. 14]{Shaw81}.

A common thread in these works is quantifying the flow of information.  To facilitate our discussion, let's first consider an intuitive definition: Information flow from process $\TSx$ to process $\TSy$ is the existence of information that is \emph{currently} in $\TSy$, the ``cause'' of which can be \emph{solely} attributed to $\TSx$'s \emph{past}. If information can be solely attributed in such a manner, we refer to it as \emph{localized}. This notion of localized flow mirrors the intuitive general definitions of ``causal'' flow proposed by Granger \cite{Granger69a} and,
before that, Wiener \cite{Wien56a}.

Ostensibly to measure information flow---and notably much later than the above efforts---Schreiber introduced the transfer entropy~\cite{schreiber2000measuring} as the information shared between $\TSx$'s past and the present $\TSy[t]$, conditioning on information from $\TSy$'s past. Perhaps not surprisingly, given the broad and pressing need to probe the organization of modern life's increasingly complex systems, the transfer entropy's use has been substantial---over the last decade and a half, its introduction alone garnered an average of $100$ citations per year.

The primary goal of the following is to show that the transfer entropy does not, in fact, measure information flow, specifically in that it attributes an information source to influences that are not localizable and so not flows. We draw out the interpretational errors, some quite subtle, that
result---including overestimating flow, underestimating influence, and more generally misidentifying structure when modeling complex systems as networks with edges given by transfer entropies.

Identifying shortcomings in the transfer entropy is not new. Smirnov~\cite{smirnov2013spurious} pointed out three: Two relate to how it responds to using undersampled empirical distributions and are therefore not conceptual issues with the measure. The third, however, was its inability to differentiate indirect influences from direct influences. This weakness motivated Sun and Bollt to propose the causation entropy~\cite{sun2014causation}. While their measure does allow differentiating between direct and indirect effects via the addition of a third hidden variable, it too ascribes an information source to unlocalizable influences.

Our exposition
reviews the notation and information theory needed and then
considers two rather similar
examples---one involving influences between two processes and the other,
influences among three. They make operational what we mean by ``localized'',
``flow'', and ``influence'', leading to the conclusion that
the transfer entropy fails to capture information flow. We
close by discussing a distinctive philosophy underlying our critique and then turn to possible resolutions and to concerns about modeling practice in network science.

\paragraph*{Background}
Following standard notation~\cite{cover2012elements}, we denote random variables with capital letters and their associated outcomes using lower case. For example, the observation of a coin flip might be denoted $\TSx$, while the coin actually landing Heads or Tails would be $\tsx$. Emphasizing temporal processes, we subscript a random variable with a time index; \eg the random variable representing a coin flip at time $t$ is denoted $\TSx[t]$. We denote a temporally contiguous block of random variables (a time series) using a Python-slice-like notation $\TSx[i][j] = \TSx[i] \TSx[i+1] \ldots \TSx[j-1]$, where the final index is exclusive. When $\TSx[t]$ is distributed according to $\Pr(\TSx[t])$, we denote this as $\TSx[t] \sim \Pr(\TSx[t])$. We assume familiarity with basic information measures, specifically the Shannon entropy $\H{\TSx}$, mutual information $\I{\TSx : \TSy}$, and their conditional forms $\H{\TSx \mid \TSz}$ and $\I{\TSx : \TSy \mid \TSz}$ \cite{cover2012elements}.

The \emph{transfer entropy} \TE{\TSx}{\TSy} from time series $\TSx$ to time series $\TSy$ is the information shared between $\TSx$'s past and $\TSy$'s present, given knowledge of $\TSy$'s past~\cite{schreiber2000measuring}:
\begin{align}
  \TE{\TSx}{\TSy} &= \I{\TSy[t] : \TSx[0][t] \mid \TSy[0][t]}
  ~.
\end{align}
Intuitively, this quantifies how much better one predicts $\TSy[t]$ using both $\TSx[0][t]$ and $\TSy[0][t]$ over using $\TSy[0][t]$ alone. A nonzero value of the transfer entropy certainly implies a kind of influence of $\TSx$ on $\TSy$. Our questions are: Is this influence necessarily via information flow? Is it necessarily direct?

Addressing the last question, the \emph{causation entropy} $\CE{\TSx}{\TSy}{\TSy,\TSz}$ is similar to the transfer entropy, but conditions on the past of a third (or more) time series \cite{sun2014causation}:
\begin{align}
  \CE{\TSx}{\TSy}{\TSy,\TSz} &= \I{\TSy[t] : \TSx[0][t] \mid \TSy[0][t], \TSz[0][t]}
  ~.
\end{align}
(It is also known as the \emph{conditional transfer entropy}.)
The primary improvement over $\TE{\TSx}{\TSy}$ is the causation entropy's ability to determine if a dependency is indirect (\ie mediated by the third process $\TSz$) or not. Consider, for example, the following system $\TSx \to \TSz \to \TSy$: variable $\TSx$ influences $\TSz$ and $\TSz$ in turn influences $\TSy$. Here, any influence that $\TSx$ has on $\TSy$ must pass through $\TSz$. In this case, the transfer entropy $\TE{\TSx}{\TSy} > \SI{0}{\bit}$ even though $\TSx$ does not directly influence $\TSy$. The causation entropy $\CE{\TSx}{\TSy}{\TSy,\TSz} = \SI{0}{\bit}$, however, due to conditioning on $\TSz$.

Many concerns and pitfalls in applying information measures comes not in their
definition, estimation, or derivation of associated properties. Rather, many
arise in \emph{interpreting} results. Properly interpreting the meaning of a
measure can be the most subtle and important task we face when using measures
to analyze a system's structure, as we will now demonstrate. Furthermore, while
these examples may seem pathological, they were chosen for their transparency
and simplicity; similar failures arise in Gaussian
systems~\cite{barrett2015exploration} signifying that the issue at hand is
widespread.

\paragraph*{Example: Two Time Series}
Consider two time series, say $\TSx$ and $\TSy$, given by the probability laws:
\begin{align*}
  \TSx[t] &\thicksim \faircoin ~, \\
  \TSy[0] &\thicksim \faircoin ~, ~\text{and} \\
  \TSy[t] &= \TSx[t-1] \oplus \TSy[t-1]
  ~;
\end{align*}
that is, $\TSx[t]$ and $\TSy[0]$ are independent and take values 0 and 1 with equal probability, and $\tsy[t]$ is the \emph{Exclusive OR} of the prior values $\tsx[t-1]$ and $\tsy[t-1]$. By a straightforward calculation we find that $\TE{\TSx}{\TSy} = \SI{1}{\bit}$. Does this mean that one bit of information is being \emph{transferred} from $\TSx$ to $\TSy$ at each time step? Let's take a closer look.

We first observe that the amount of information in $\TSy[t]$ is $\H{\TSy[t]} = \SI{1}{\bit}$. Therefore, the uncertainty in $\TSy[t]$ can be reduced by at most \SI{1}{\bit}. Furthermore, the information shared by $\TSy[t]$ and the prior behavior of the two time series is $\I{\TSy[t] : \left( \TSx[0][t], \TSy[0][t] \right)} = \SI{1}{\bit}$. And so, the \SI{1}{\bit} of $\TSy[t]$'s uncertainty in fact can be removed by the prior observations of both time series.

\begin{figure}
  \centering
  \subfloat[{\protect $\TSy[0][t]$} alone does not predict {\protect $\TSy[t]$}. (The {\protect \drawmia}-shaped region {\protect $\I{\TSy[0][t] : \TSy[t]} = \SI{0}{\bit}$}.) However, when used in conjunction with {\protect $\TSx[0][t]$}, they completely predict its value. (The {\protect \drawtea}-shaped region {\protect $\I{\TSx[0][t] : \TSy[t] \mid \TSy[0][t]} = \SI{1}{\bit}$}.)]
  {
    \includegraphics{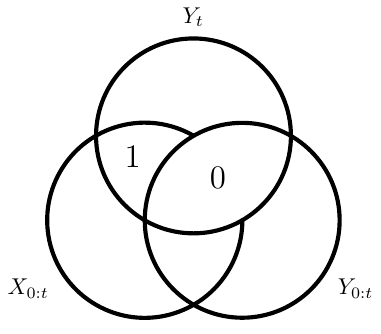}
    \label{subfig:TE_one}
  }%
  \subfloat[$\TSx$'s past {\protect$\TSx[0][t]$} alone does not aid in predicting {\protect$\TSy[t]$}. (The {\protect \drawmib}-shaped region {\protect $\I{\TSx[0][t] : \TSy[t]} = \SI{0}{\bit}$}.) However, given knowledge of {\protect$\TSx[0][t]$}, then {\protect$\TSy[0][t]$} can predict {\protect$\TSy[t]$}. (The {\protect \drawteb}-shaped region {\protect $\I{\TSy[0][t] : \TSy[t] \mid \TSx[0][t]} = \SI{1}{\bit}$}.)]
  {
    \includegraphics{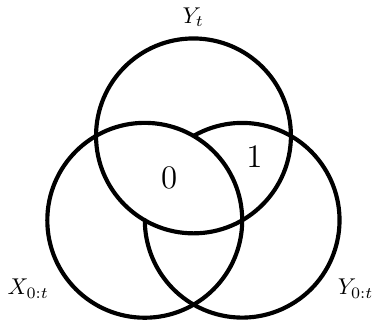}
    \label{subfig:TE_two}
  }
  \caption{Two complementary ways to view the information shared between
  $\TSx[0][t], \TSy[0][t]$, and $\TSy[t]$. In each I-Diagram, a circle
  represents a random variable whose area measures the random variable's
  entropy. Overlapping regions are information that is shared. The transfer
  entropy is a conditional mutual information; a region where two random
  variables overlap, but that falls outside the random variable being
  conditioned on.}
  \label{fig:transfer_entropy}
\end{figure}

How much does $\TSy[0][t]$ alone help us predict $\TSy[t]$? We quantify this using mutual information. Since $\I{\TSy[t] : \TSy[0][t]} = \SI{0}{\bit}$, the variables are independent: $\TSy[0][t]$ alone does not help in predicting $\TSy[t]$. However, knowing $\TSy[0][t]$, how much does $\TSx[0][t]$ help in predicting $\TSy[t]$? The conditional mutual information $\I{\TSy[t] : \TSx[0][t] \mid \TSy[0][t]} = \SI{1}{\bit}$---the transfer entropy we just computed---quantifies this. This situation is graphically analyzed via the information diagram (I-diagram)~\cite{yeung1991new} in Fig.\nobreakspace \ref {subfig:TE_one}.

To obtain a more complete picture of the information dynamics under consideration, let's reverse the order in which the time series are queried. The mutual information $\I{\TSy[t] : \TSx[0][t]} = \SI{0}{\bit}$ tells us that the $\TSx$ time series alone does not help predict $\TSy[t]$. However, the conditional mutual information $\I{\TSy[t] : \TSy[0][t] \mid \TSx[0][t]} = \SI{1}{\bit}$. And so, from this point of view it is $\TSy$'s past that helps predict $\TSy[t]$, contradicting the preceding analysis. This complementary situation is presented diagrammatically in Fig.\nobreakspace \ref {subfig:TE_two}.

How can we rectify the seemingly inconsistent conclusions drawn by these two
lines of reasoning? The answer is quite straightforward: the \SI{1}{\bit} of information about $\TSy[t]$ does not come from \emph{either} time series individually, but rather from \emph{both} of them simultaneously.
(In fact, the I-Diagrams are naturally consistent, once one recognizes that the \emph{co-information}~\cite{Bell03a}, the inner-most information atom, is $\I{\TSy[t] : \TSx[0][t] : \TSy[0][t]} = \SI{-1}{\bit}$.)

In short, the \SI{1}{\bit} of reduction in uncertainty $\H{\TSy[t]}$ should not be \emph{localized} to either time series. The transfer entropy, however, erroneously localizes this information to $\TSx[0][t]$. In light of this, the transfer entropy \emph{overestimates} information flow.

This example shows that the transfer entropy can be positive due not to
information flow, but rather to nonlocalizable influence---in this case, a
\emph{conditional dependence} between variables. This suggests that, though inappropriate for measuring information flow, the transfer entropy may be a viable measure of such influence. Our next example illustrates that this too is incorrect.

\paragraph*{Example: Three Time Series}
Our second example parallels the first. Before, we considered the case where \emph{one} of two time series is determined by the past of \emph{both}, we now consider the case where \emph{two} time series determine a \emph{third}, again via an Exclusive OR operation. Their probability laws are:
\begin{align*}
  \TSx[t] &\thicksim \faircoin ~,\\
  \TSy[t] &\thicksim \faircoin ~,~\text{and}\\
  \TSz[t] &= \TSx[t-1] \oplus \TSy[t-1]
  ~,
\end{align*}
in which $\tsz[0]$'s value is irrelevant. Unlike the prior example,
the transfer entropy from either $\TSx$ or $\TSy$ to
$\TSz$ is zero: $\TE{\TSx}{\TSz} = \TE{\TSy}{\TSz} = \SI{0}{\bit}$, and it
therefore \emph{underestimates} influence that is present. Furthermore, the
relevant pairwise mutual informations all vanish: $\I{\TSz[t] : \TSx[0][t]} = \I{\TSz[t] : \TSy[0][t]} = \I{\TSz[t] : \TSz[0][t]} = \SI{0}{\bit}$. The time series are pairwise independent.

\begin{figure}
  \centering
  \contourlength{2pt}
  \includegraphics{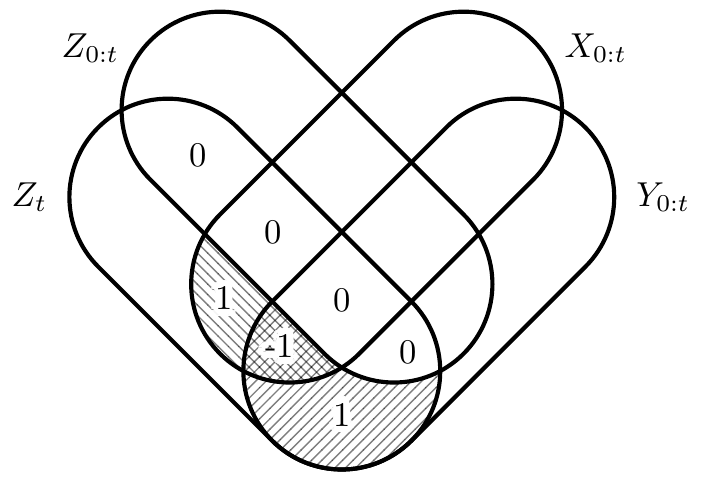}
  \caption{Information diagram depicting both transfer entropies and causation entropies for three times series $\TSx$, $\TSy$, and $\TSz$. $\TE{\TSx}{\TSz} = \SI{0}{\bit}$ corresponds to the two regions shaded with south-east sloping lines and $\TE{\TSy}{\TSz} = \SI{0}{\bit}$, the two regions shaded with north-east sloping lines. $\CE{\TSx}{\TSz}{\TSy,\TSz} = \SI{1}{\bit}$ is the region containing only south-east sloping lines and, similarly, $\CE{\TSy}{\TSz}{\TSx,\TSz} = \SI{1}{\bit}$ is the region containing only north-east sloping lines.}
\label{fig:causation_entropy}
\end{figure}

Given that we are probing the influences between three time series, it is natural now to consider the behavior of the causation entropy. In this case, we have $\CE{\TSx}{\TSz}{\TSy,\TSz} = \CE{\TSy}{\TSz}{\TSx,\TSz} = \SI{1}{\bit}$, indicating that given the past behavior of $\TSz$ and $\TSx$ (or $\TSy$), the past of $\TSy$ (or $\TSx$) can be used to predict the behavior of $\TSz[t]$. Like before, this \SI{1}{\bit} of information cannot be localized to either $\TSx$ or $\TSy$ and so it is inaccurate to ascribe the \SI{1}{\bit} of information in $\TSz[t]$ to either $\TSx$ or $\TSy$ alone. In this way, the causation entropy also erroneously localizes the \SI{1}{\bit} of joint influence. While the causation entropy succeeds here as a measure of nonlocalizable influence, as a measure of information flow, it overestimates. (This is known to Sun and Bollt, but here we stress that the failure is a general issue with interpreting its value, not merely a limitation regarding network inference.) These information quantities are displayed in the I-Diagram in Fig.\nobreakspace \ref {fig:causation_entropy}.

\paragraph*{Discussion}
We see that transfer-like entropies can both overestimate information flow
(first example) and underestimate influence (second example). The primary misunderstanding of these quantities stems from a mischaracterization of the conditional mutual information. Most basically, probabilistic conditioning is not a ``subtractive'' operation: \I{\TSx : \TSy \mid \TSz} is not the information shared by $\TSx$ and $\TSy$ once the influences of $\TSz$ have been removed. Rather, it is the information shared by $\TSx$ and $\TSy$ \emph{taking into account} $\TSz$. This is not a game of mere semantics: Conditioning can \emph{increase} the information shared between two processes: $\I{\TSx : \TSy} < \I{\TSx : \TSy \mid \TSz}$. This cannot happen if conditioning merely removed influence: conditional dependence includes \emph{additional} dependence that occurs in the presence of a third variable~\cite{nemenman2004information}. Measuring information flow---as we have defined it---requires a method of \emph{localizing} information. Since simple conditioning can fail to localize information, the transfer entropy, causation entropy, and other measures utilizing the conditional mutual information can fail as measures of information flow.

Another way to understand conditional dependence is through the \emph{partial information decomposition}~\cite{williams2010nonnegative}. Within this framework, the mutual information between two random variables $X_1$ and $X_2$ (call them \emph{inputs}) and a third random variable $Y$ (the \emph{output}) is decomposed into four mutually exclusive components: $\I{(X_1, X_2) : Y} = R + U_1 + U_2 + S$. $R$ quantifies how the inputs $X_1$ and $X_2$ \emph{redundantly} inform the output $Y$, $U_1$ and $U_2$ quantify how each provides \emph{unique} information to $Y$, and finally $S$ quantifies how the inputs together \emph{synergistically} inform the output. In this decomposition, the mutual information between one input and the output is equal to what uniquely comes from that input plus what is redundantly provided by both inputs; $\I{X_1 : Y} = R + U_1$, for example. However, the mutual information between that input and the output conditioned on the other input is equal to what uniquely comes from that one input, plus what is synergistically provided by both inputs: $\I{X_1 : Y \mid X_2} = U_1 + S$. In other words, conditioning removes the redundant information, but adds the synergistic information. Here, conditional dependencies manifest themselves as synergy.

Treating $\TSx[0][t]$ and $\TSy[0][t]$ as inputs and $\TSy[t]$ as output, the
partial information decomposition identifies the transfer entropy
\TE{\TSx}{\TSy} as the sum of the unique information from $\TSx[0][t]$ plus the
synergistic information from both $\TSx[0][t]$ and $\TSy[0][t]$ together. It
seems natural, and has been previously
proposed~\cite{williams2011generalized,barrett2015exploration}, to associate
only this unique information with information flow. The transfer entropy,
however, conflates unique information and synergistic information leading to
inconsistencies, such as analyzed in the examples. Similar conclusions follow
for the causation entropy; however, due to the additional variable, the
analysis is more involved.

Though there is as yet no broadly accepted quantification of unique
information~\cite{bertschinger2014quantifying}, if one were able to accurately
measure it, it may prove to be a viable measure of information flow. It is
notable that Stramaglia \etal, building on
Ref.~\cite{bettencourt2008identification}, considered how total synergy and
redundancy of a collection of variables influence
each other~\cite{stramaglia2012expanding}.

Other quantifications of information flow between time series have been
proposed. The \emph{directed information}~\cite{massey1990causality} is
essentially a sum of transfer entropies and so inherits the same flaws.
Furthermore, both the transfer entropy and directed information have been shown
to be generalizations of \emph{Granger causality}~\cite{Granger69a,
barnett2009granger,amblard2011directed,Footnote1}, itself purportedly a measure
of ``predictive causality''~\cite{diebold2007elements}. Ay and Polani proposed
a measure of information flow based on active intervention in which an outside
agent modifies the system in question by removing
components~\cite{ay2008information}. We conjecture that all these measures
suffer for the same reasons---conflation of dyadic and polyadic relationships.

\paragraph*{Conclusions and Consequences}
Although the examples were intentionally straightforward, the consequences
appear far-reaching. Let's consider network
science~\cite{newman2003structure} which, over the same decade and a half
period since the introduction of the transfer entropy, has developed into a
vast and vibrant field, with significant successes in many application areas.
Standard (graph-based) networks are composed of \emph{nodes}, representing
system observables, and \emph{edges}, representing relationships between them.
As commonly practiced, such networks represent dyadic (binary) relationships
between nodes~\cite{Footnote2}---article co-authorship, power transmission
between substations, and the like. It would seem, then, that much of the
popularity of using the transfer entropy to analyze large-scale complex systems
is that it is an information measure specifically adapted to quantifying dyadic
relationships. Such a tool goes hand-in-hand with standard network modeling.

As the examples emphasized, though, observables may be related by polyadic
relationships that cannot be naturally represented on a standard network as
commonly practiced. For example, all three variables in our second example are
pairwise independent. A standard network representing dependence between them
therefore consists of three disconnected nodes, thus failing to capture the
dependence between variables that is, in fact, present. As a start to repair
this deficit, it would be more appropriate to represent such a complex system
as a \emph{hypergraph}~\cite{Ramanathan11a,Estr05a}.

Continuing this line of thought, if one believes that a standard network is an accurate model of a complex system, then one implicitly assumes that polyadic relationships are either not important or do not exist. Said this way, it is clear that when modeling a complex system, one must test for this lack of polyadic relationship first. With this assumption generally unspoken, though, it is not surprising that a nonzero value of the transfer entropy leads analysts to interpret it as information flow. Within that narrow view, indeed, how else could one time series influence another if all interactions are dyadic? Restated, when a system is modeled as a standard network, all relationships are assumed to be dyadic. One is therefore naturally inclined to explain all observed dependencies as being dyadic. The cost, of course, is either a greatly impoverished or a spuriously embellished view of organization in the world. As such, modeling a complex system by way of a graph with edges determined by transfer or causation entropies is intrinsically flawed.

Many of the preceding issues are difficult to analyze since at present notions
of ``influence'' are not sufficiently precise and, even when they are as with
the use of information diagrams and measures and the partial information
decomposition, there is a combinatorial explosion in possible types of
dependence relationships. Said differently, what one needs is a more explicit,
even more elementary, structural view of how one process can be transformed to
another. Paralleling the canonical \eM minimal sufficient statistic
representation of stationary processes, two of us (NB and JPC) recently
introduced a minimal optimal transformation of one process into another, the
\eT~\cite{Barn13a}. This provides a structural representation for the minimal
optimal predictor of one process about another. The corresponding transducer analysis,
paralleling that above in Figs.\nobreakspace \ref {fig:transfer_entropy} and\nobreakspace  \ref {fig:causation_entropy},
identifies new informational atoms beyond those of the transfer entropies~\cite{Barn14a}.

In short, the transfer entropy can both overestimate information flow (first
example) and underestimate influence (second example). These effects are
compounded when viewing complex systems as standard networks since the latter
further misconstrue polyadic relationships. While we do not object to the
transfer entropy as a measure of the reduction in uncertainty about one time
series given another, we do find its mechanistic interpretation as information \emph{flow}
or \emph{transfer} to be incorrect. In fact, this is true for any related
measures---such as the causation entropy---that are based on conditional mutual
information between observed variables. In light of these interpretational
concerns, it seems that several recent works that rely heavily on transfer-like
entropies---ranging from cellular automata~\cite{lizier2008local} and
information thermodynamics~\cite{parrondo2015thermodynamics} to cell regulatory
networks~\cite{walker2015informational} and
consciousness~\cite{lee2015assessing}---will benefit from a close reexamination.

We thank A. Boyd, K. Burke, J. Emenheiser, B. Johnson, J. Mahoney, A.
Mullokandov, P.-A. No{\"e}l, P. Riechers, N. Timme, D. P. Varn, and G. Wimsatt
for helpful feedback. This material is based upon work supported by, or in part
by, the U. S. Army Research Laboratory and the U. S. Army Research Office under
contracts W911NF-13-1-0390 and W911NF-13-1-0340.



\begin{thebibliography}{10}
\bibitem{Footnote1}
Granger causality can refer to either Granger's general intuitive definition of
  predictive causality or the specific (linear) statistical methods that he
  proposed. We refer to the latter, more commonly used meaning.
\bibitem{Footnote2}
Higher-order dependencies can be represented with standard networks using
  additional, so-called latent variables. For example, one can represent
  polyadic relationships by building a new bipartite network consisting of the
  original nodes (type A) plus additional nodes representing polyadic
  relationships (type B). Here, an edge exists between a node of type A and a
  node of type B if that node is involved in that polyadic relationship. In any
  case, directly measuring and interpreting information flow between nodes
  becomes a much more subtle issue in such augmented, hidden-variable networks.
\end{thebibliography}
\end{document}